\documentclass[11pt,twoside]{article}
\usepackage{./asp2014}

\aspSuppressVolSlug
\resetcounters

\begin{document}

\title{The Ultimate Display}
\author{C.J.~Fluke$^{1,2}$ and D.G.~Barnes,$^2$
\affil{$^1$Swinburne University of Technology, Hawthorn, Victoria, Australia; \email{cfluke@swin.edu.au}}
\affil{$^2$Monash University, Clayton, Victoria, Australia;}}

\paperauthor{Christopher~Fluke}{cfluke@swin.edu.au}{}{Swinburne University of Technology}{Centre for Astrophysics \& Supercomputing}{Hawthorn}{Victoria}{3122}{Australia}
\paperauthor{David~Barnes}{david.g.barnes@monash.edu}{}{Monash University}{Monash eResearch Centre}{Clayton}{Victoria}{3168}{Australia}

\begin{abstract}
Astronomical images and datasets are increasingly high-resolution and
multi-dimensional.  The vast majority of astronomers perform all of their
visualisation and analysis tasks on low-resolution, two-dimensional
desktop monitors.   If there were no technological barriers to designing
the ultimate stereoscopic display for astronomy, what would it look
like?  What capabilities would we require of our compute hardware to
drive it?  And are existing technologies even close to providing a true
3D experience that is compatible with the depth resolution of human
stereoscopic vision?  We consider the CAVE2 (an 80 Megapixel, hybrid 2D and
3D virtual reality environment directly integrated with a 100 Tflop/s
GPU-powered supercomputer) and the Oculus Rift (a low-cost, head-mounted
display) as examples at opposite financial ends of the immersive display
spectrum. 
\end{abstract}
\section{Introduction}
Astronomical datasets, from observational programs and numerical simulations, are complex, multi-dimensional, high-resolution, and with a high dynamic range.  The displays used by most astronomers are simple, two-dimensional, low-resolution, with 24-bit colour.  A great deal of effort is invested in gathering photons and generating bits and bytes -- but are we making the best use of the astronomer's personal visual processing system to discover knowledge?  One solution is to place the astronomer in a display environment where all that can be seen is data.

We concentrate on opportunities to make better use of immersive stereoscopic displays, where the astronomer's attention is focussed entirely on the three-dimensional features of the data.  Stereoscopic (or stereo) imaging requires the creation of a pair of images, one for each eye.  These images are directed to the correct eye using, for example, glasses (polarising, electronic shutters, two-colour anaglyph, etc.), glasses-free parallax barriers, or a unique display screen for each eye.  

\articlefiguretwo{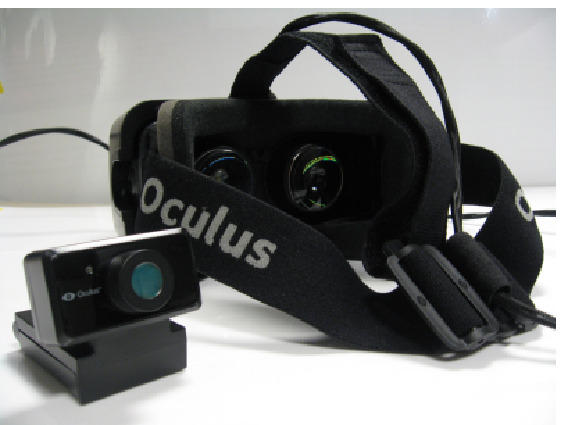}{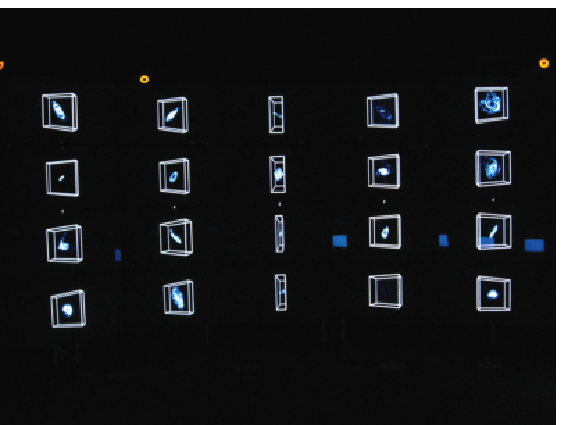}{fig:ids}{\emph{(Left:)} An Oculus Rift DK2 head-mounted display.  \emph{(Right:)} 
A portion of the CAVE2 at Monash University.}

The Ultimate Display\footnote{We make no claim to be the first to consider this matter, and acknowledge that Virtual Reality pioneer, Ivan \citet{Sutherland65}, wrote {\em the} landmark work on virtual environments under the same title} has three main requirements.   Firstly, it should fill the viewer's visual field-of-view (FOV), thus minimising distraction from other events.   This is typically quoted as $\sim\!\!150^{\circ}$ horizontal \citep{West94} and $\sim\!\!\!120^{\circ}$ vertical per eye.  Due to the occluding effects of the viewer's nose, the horizontal overlap region between eyes is closer to $100-120^{\circ}$.   
For simplicity, we choose $110^{\circ}$ in both directions.  

Secondly, it should use a pixel size (at least) as small as the eye can resolve. Visual acuity ($V_{\rm A}$) is the ability to spatially resolve adjacent pixels.  Ignoring wavelength dependence and direction-based variable resolution, $V_{\rm A}$ is $\sim\!30$ arcsec \citep[e.g.][]{West94,Deering98}. The limiting factors are diffraction effects in the human optical system and the spacing of cones within the fovea \citep{West12}. 

Finally, the display should reach the binocular disparity limit.  Stereo acuity is an example of hyperacuity, where optical and anatomical limits to resolution are overcome by neural processing.  By combining two independent images (the stereo pair), subtle spatial differences are processed and identified within the viewer's brain (Westheimer 1994; 2012).   Stereo acuity may be as low as a few arcseconds, or almost an order of magnitude better than visual acuity for some astronomers.

\section{Technology}
To determine whether current technologies are compatible with the Ultimate Display,  we consider two alternatives with vastly different price points: head-mounted displays (HMDs) such as the Oculus Rift
\footnote{\url{http://www.oculus.com}}
 and large-format, hybrid visualisation and supercomputing environments typified by the CAVE2
\footnote{\url{https://www.evl.uic.edu/cave2}}
 (see Figure \ref{fig:ids}).

As the name suggests, HMDs are personal devices worn over the head with a dedicated screen for each eye.  Utilising motion and position tracking, the viewer is given a a  window into a $4\pi$ steradian virtual world.       
The Oculus Rift is one of a growing number of low-cost, commodity HMDs, including products such as the Samsung Gear VR and Sony's Project Morpheus.   
The Rift DK2 offers $960 \times 1080$ pixels on a 3.3-inch screen per eye.   The FOV is $\sim\!100^{\circ}$ horizontal and vertical for each eye.  In-built lenses focus the images, and software-corrections overcome chromatic aberration.

Developed at the Electronic Visualization Laboratory (University of Illinois at Chicago),  the CAVE2 combines the best features of the traditional CAVE (i.e. multi-wall, rear projection, stereoscopic) and tiled display walls \citep[TDWs; see][]{Meade14}.  Wrapping the TDW around the viewer replicates the CAVE's ``rear projection'' display mode (but with a visual impact due to the bezels). Replacing the individual display components with stereoscopic panels provides the immersive stereo environment.  The total number of pixels, and the overall display brightness, increase significantly compared to projection.   Furthermore, the graphics card capabilities (memory and fillrate) required to generate and display stereo content also increase.  

The CAVE2 at Monash University (Australia) comprises twenty columns and four rows of monitors.  Each element is a 46-inch (diagonal) $1366 \times 768$ Planar Matrix LX46L 3D monitor (34 pixels per inch -- see below).  This gives 80 million pixels in 2D mode or 40 million (per eye) for stereo via row interlacing.  The Monash CAVE2 is 8 metres in diameter, with a $30^{\circ}$ gap for ingress.  Each column is driven by a 1536-core NVIDIA K5200 graphics card.  A second GPU/column is available for computation, making this a hybrid supercomputer/display with 
$\sim\!\!100$ Tflop/s processing power.

From a viewing distance, $d$ (metres), 
with an angular pixel size, $\theta_{\rm p}$, the target pixels per inch, $P_{\rm PI}$, compatible with visual acuity, 
$V_{\rm A} = 30$ arcsec, is:
\begin{equation}
P_{\rm PI} \approx 174.6 (V_{\rm A}  /\theta_{\rm p}) \left( 1/d \right).
\end{equation}

Using rectangular display components with horizontal and vertical pixel resolutions $P_{\rm H}$ and $P_{\rm V}$,
the horizontal dimension, $H_{\rm D}$, required to achieve the target $P_{\rm PI}$ is:
\begin{equation}
H_{\rm D} = P_{\rm H}/P_{\rm PI} \hspace{0.2 cm} \mbox{inches} \hspace{0.2 cm} = 10.997 \left(P_{\rm H}/1920\right) 
(\theta_{\rm p}/V_{\rm A}) d \hspace{0.2 cm} \mbox{inches} .
\end{equation}

For a cylindrical display configuration to fill the horizontal FOV, $\theta_{\rm H}$ (degrees), the number of tiled display elements is:
\begin{equation}
n_{\rm H} = \left\lceil{\frac{C}{H_D}}\right\rceil \approx \left\lceil{   6.9 \left(\frac{\theta_{\rm H}}{110^\circ}\right) 
\left(\frac{1920}{P_H}\right)  \left(\frac{V_{\rm A} }{\theta_{\rm p}}\right)} \right\rceil 
\hspace{0.2cm} \mbox{with} \hspace{0.2cm} C = \theta_{\rm H}  \left( \frac{39.37 \pi}{180} \right)d  \hspace{0.2cm} \mbox{inches}.
\end{equation}
A weaker limit for tiling the vertical FOV is: $n_{\rm V} \lesssim n_{\rm H} \left(P_{\rm H}/P_{\rm V}\right)$.

Table \ref{tbl:results} encapsulates the results for FullHD ($1920 \times 1080$), 4K ($3840 \times 2160$) and 8K ($7680 \times 4320$) consumer-style display elements.  We use [**] to indicate configurations that are achievable with current or near-term  technology.  For the CAVE2 approach, we need to relax the $n_{\rm V}$ constraint for larger viewing distances -- unless we suspend the viewer within the environment -- as the vertical FOV intersects the ground.    Using more displays with smaller $H_{\rm D}$ is deemed an acceptable solution: this reduces $P_{\rm PI}$ below visual acuity but strengthens stereo immersion.

\begin{table}
\caption{Target pixels per inch, $P_{\rm PI}$, to match visual acuity ($V_{\rm A}  = \theta_{\rm P} = 30$ arcsec). We assume $\theta_{\rm H} = \theta_{\rm V} = 110^{\circ}$ for the FOV.  For different pixel dimensions ($P_{\rm H} \times P_{\rm V}$), we calculate the horizontal, $H_{\rm D}$ and diagonal, $D_{\rm D}$, dimensions (inches) and the number, $n_{\rm H}$ and $n_{\rm V}$, to fill the FOV.  [**] indicates reasonable (near-future) solutions -- 
some of the $D_{\rm D}$ values are optimistic.}
\label{tbl:results}
\begin{center}
\begin{tabular}{lccccccccc}
{\bf Device} & $d$ (m)  & $P_{\rm PI}$ &  $P_{\rm H} \times P_{\rm V}$ & $H_{\rm D}$ & $D_{\rm D}$ & $n_{\rm H}$ & $n_{\rm V}$ & \\ 
\hline
HMD-style & 0.05 &  3500 &1920 $\times$ 1080 &   0.55 & 0.63&  7 & 13 \\
& & &  3840 $\times$ 2160 & 1.1& 1.26 & 4 & 8 \\
& & & 7680 $\times$ 4320& 2.2 & 2.52 & 2 & 4 \\
CAVE2-style & 5.0  &35 &1920 $\times$ 1080 &   55 & 63 & 7 & 13 \\
& & &  3840 $\times$ 2160 & 110 & 126 & 4 & 8 & [**] \\
& & &  7680 $\times$ 4320& 220 & 252 & 2  & 4 & \\
&&& 1366 $\times$ 768 & 39 & 45 & 10 & 18 &[**] \\
\hline
\end{tabular}
\end{center}
\end{table}

Now consider the binocular disparity limit, where $P_{\rm PI}$ required for true stereo immersion will be $\sim\!\!\!10$ times the values in Table 
\ref{tbl:results}.  Whereas a 4x8 tiled display using 120-inch 4K screens viewed at 5 m is a close match to visual acuity, this is only providing one tenth of the potential depth-based information.   No wonder, then, that existing desktop-based stereo displays and HMDs provide only minimal depth clarity.  This tallies with our direct experience of immersive visualisation in the CAVE2 compared to lower-pixel count flat screens (desktop and large-format) and HMDs.  CAVE2 stereo {\em is} more compelling, despite the bezel-induced image discontinuities.  To increase the number of horizontal pixels for the 2x4 tile, 252-inch 8K configuration, we would need 10 x 25-inch 8K screens, which would encompass the factor of 2 increase in pixels required vertically for row interlacing.   An upgrade path might be to swap out each existing panel with a (future) 8K row-interlaced panel.

Up to this point, we have concentrated on the pixels themselves - not on the graphics hardware required to actually fill the desired number of pixels at a reasonable frame rate (Deering 1998).   At the binocular disparity limit, we need to generate $\sim\!\!10 P_{\rm H} P_{\rm V} n_{\rm h} n_{\rm v}$ pixels per frame (total), which can be solved with multiple graphics cards.   The Monash CAVE2 currently renders 80 Megapixels with twenty K5200 cards (20 Gigapixel/s fillrate per card gives 400 Gigapixels/s total or 5 Gigapixel/s fillrate per Megapixel on screen).   The Ultimate CAVE2 with a few Gigapixels, requiring 5000 Gigapixel/s fillrate (total), is plausible today with 50 NVIDIA Quadro M6000 cards.

\section{Concluding Remarks}
As a low-cost, highly portable solution, HMDs remain an intriguing option.  From the perspective of display quality, HMDs clearly have a long way to go before they match either the visual acuity or binocular disparity limit.  It seems very unlikely that there will be orders of magnitude changes here in the short term.

A CAVE2 configuration can reach the visual acuity limit and fill the horizontal FOV, but does not fare quite so well with vertical coverage.   Reaching the binocular disparity limit, and having the graphics capability to display content at a reasonable frame-rate, is plausible in the near-term.  We acknowledge that the CAVE2 is a high-end, expensive solution, unlikely to be installed at all research institutions.  However, with large collaborations, it is not necessary for every team member to be directly involved in visualisation-based knowledge discovery.  
The CAVE2 appears to be on an achievable trajectory for astronomers to have their Ultimate Display.

\bibliography{O10-4}  

\end{document}